\documentclass{article}

\usepackage{amsfonts, amsmath, amssymb, amsbsy}
\usepackage{epsfig}
\usepackage{graphicx}
\usepackage{longtable}

\begin{document}

\newcommand{\msun}{\,{M}_{\odot}}
\newcommand{\lsun}{\,{L}_{\odot} }
\newcommand{\mJy}{\,\rm mJy}
\newcommand{\ergs}{\,{\rm erg\,s}^{-1}}
\newcommand{\esc}{\,{\rm erg\,s^{-1}\,cm^{-2}}}
\newcommand{\kms}{\,{\rm km\,s^{-1}}}      
\newcommand{\mum}{\,\mu{\rm m}}
\newcommand{\mbh}{$M_{\rm BH}$}
\newcommand{\mmbh}{M_{\rm BH}}
\newcommand{\lbol}{$L_{\rm bol}$}
\newcommand{\mlbol}{L_{\rm bol}}
\newcommand{\rf}{\par\noindent\hangindent 15pt {}}
\newcommand{\solar}{L$_{\odot}$\ }
\newcommand{\solm}{M$_{\odot}$\ }

\begin{center}
\textbf{\large The Center of the Milky Way from Radio to X-rays}
\end{center}

\begin{center}
\textbf{
 A. Eckart$^{1,2}$, 
 M. Valencia-S.$^{1}$, 
 B. Shahzamanian$^{1,2}$, 
 M. Garc\'{i}a-Mar\'{i}n$^{1}$, 
 F. Peissker$^{1}$, 
 M. Zajacek$^{1,2,3}$, 
 M. Parsa$^{1,2}$, 
 B. Jalali$^{1}$, 
 R. Saalfeld$^{1}$, 
 N. Sabha$^{1.}$, 
 S. Yazici$^{1}$,
 G. D. Karssen$^{1}$, 
 A. Borkar$^{1,2}$, 
 K. Markakis$^{1,2}$, 
 J.A. Zensus$^{2,1}$, 
 C. Straubmeier$^{1}$
}
\end{center}

\begin{center}
{\it
\noindent 
$^1$
I. Physikalisches Institut der Universit\"at zu K\"oln, Z\"ulpicher Str. 77, D-50937 K\"oln, Germany;
\\
$^2$
Max-Planck-Institut f\"ur Radioastronomie, Auf dem H\"ugel 69, D-53121 Bonn, Germany;
\\
$^3$
Astronomical Institute of the Academy of Sciences Prague, Bocni II 1401/1a, \\
CZ-141 31 Praha 4, Czech Republic
}
\end{center}

\begin{abstract}
We summarize basic observational results on Sagittarius~A* obtained from the radio, infrared and X-ray domain.
Infrared observations have revealed that a dusty S-cluster\footnote{The S-cluster is the cluster of high velocity 
stars surrounding SgrA*; see Eckart\&Genzel (1997).} object (DSO/G2) passes by SgrA* within $\sim$120~AU, 
the central super-massive black hole of the Milky Way.
It is still expected that this event will give rise to exceptionally intense activity in the 
entire electromagnetic spectrum.
Based on February to September 2014 SINFONI observations
the detection of a spatially compact and red-shifted hydrogen recombination line emission allowed us to obtain a new 
estimate of the orbital parameters of the DSO.
We have not detected strong pre-peribothron\footnote{Periapse of orbit around a black hole} blue-shifted nor post-peribothron red-shifted emission above the noise level at 
the position of SgrA* or upstream the orbit. 
The peribothron position was reached in May 2014.
Our 2004-2012 infrared polarization statistics show that SgrA* must be a very stable system - both in terms
of the geometrical orientation of a jet or accretion disk and in terms of the variability spectrum 
which must be linked to the accretion rate.
Hence, polarization and variability measurements are the ideal tools to probe for any change in the system as a 
function of the DSO/G2 fly-by.
Due to the 2014 fly-by of the DSO, an increased accretion activity of SgrA* may still be upcoming.
Future observations of bright flares will improve the derivation of the spin and
the inclination of the SMBH from NIR/sub-mm observations.

\end{abstract}

\section{Introduction}

Sagittarius~A* (Sgr~A*) at the center of our galaxy is a highly variable 
near-infrared (NIR) and X-ray source which is associated with a 
$ 4 \times 10^{6}$ \solm super-massive central black hole (SMBH).
Gillessen et al. (2012) report a fast moving
infrared excess source which they interpret as a core-less gas and dust cloud
approaching SgrA* on an elliptical orbit.
Eckart et al. (2013ab) present K$_s$-band identifications (from VLT and Keck data)
and proper motions of the DSO. 
The DSO is only one of several infrared excess sources in the central few arcsecond
(see Fig.~\ref{ident}a).
In Shahzamanian et al. (2015) 
we present new results obtained from observations of polarized near-infrared (NIR) light from Sgr A*.
The observations have been carried out using the adaptive optics
instrument NACO at the VLT UT4 in the infrared K$_s$-band (2.00$\mu$m - 2.36$\mu$m) from 2004 to 2012.
Several polarized flares were observed during these years, allowing us to study the statistical
properties of linearly polarized NIR light from Sgr A*.

\section{Results}
\subsection{Infrared Imaging Spectroscopy}

\begin{figure}[!ht]
\begin{center}
\includegraphics[width=13cm]{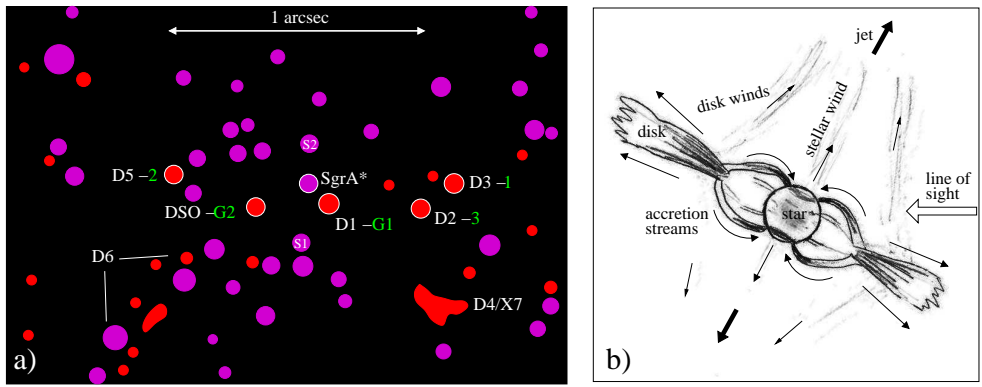}
\caption{a) Sketch of the distribution of stars and dusty objects in the central 2.4''$\times$1.6'' of the Milky Way
based on K- and L'-band images by Meyer et al. (2014) and Eckart et al. (2013).
With the exception of SgrA* we have coded the nomenclature used by Eckart et al. (2013) in white and 
the nomenclature used by Meyer et al. (2014) in green color.
For the cometary tailed dusty source X7 see also  Muzic et al. 2010.
b) Sketch of a young star - seen edge on and not to scale. 
Possible source regions for a young accreting star to generate broad lines are indicated by arrows.
} \label{ident}
\end{center}
\end{figure}

\begin{figure}[!ht]
\begin{center}
\includegraphics[width=10cm]{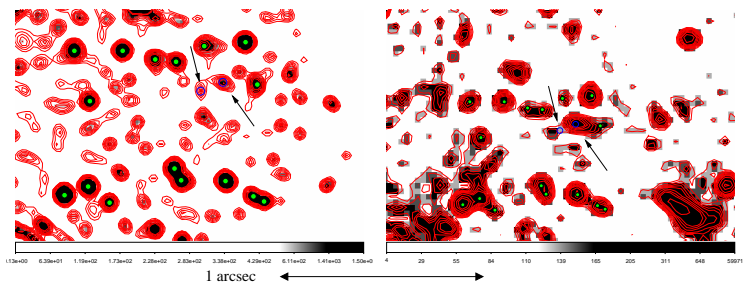}
\caption{The last 2013 (September) images of the Galactic center in the NIR K-band (left) and L'-band (right) 
before NACO was dismounted from UT4 
and the source went into the southern summer and was only observable again in early 2014.
The (confused) position of the DSO is marked by a downwards arrow and the position of SgrA* by an upward arrow.
Both positions are also marked by a blue circle.
} \label{KLimages}
\end{center}
\end{figure}

Based on L'-band imaging, an infrared excess source within the central cluster of high velocity S-stars was
found to approach the immediate vicinity of SgrA* (Gillessen et al. 2012).
In addition, Br$\gamma$ line emission was reported by Gillessen et al. (2013a) and Phifer et al. (2013). 
In Eckart et al. (2013ab) we report the identification of K$_s$-band emission from a source at the position 
of the L'-band identification. The proper motions of all accessible K$_s$-band data agree well with those obtained from
L'-band and Br$\gamma$ line emission.
In 2013 the source is confused by the presence of fore- or background sources in the L'-band and probably also in the 
K$_s$-band (Fig.~\ref{KLimages}).
Phifer et al. (2013) show that in addition to the background flux at the position of the DSO  no source brighter 
than m$_K$=20 can be determined.
Gillessen et al. (2013b) report a marginal spatial extension of the Br$\gamma$ line emission in their SINFONI data
and find an intrinsic Gaussian FWHM size of $42\pm10$~mas. Given the peculiar orientation of the source estimated 
orbit, precise determinations of the source elongation along the orbit are difficult to obtain.
Combining these observational facts indicate that a dusty object 
- possibly associated with a stellar object (see Fig.~\ref{ident}b) -
is on an elliptical orbit around SgrA*.

\begin{figure}[!ht]
\begin{center}
\includegraphics[width=9cm]{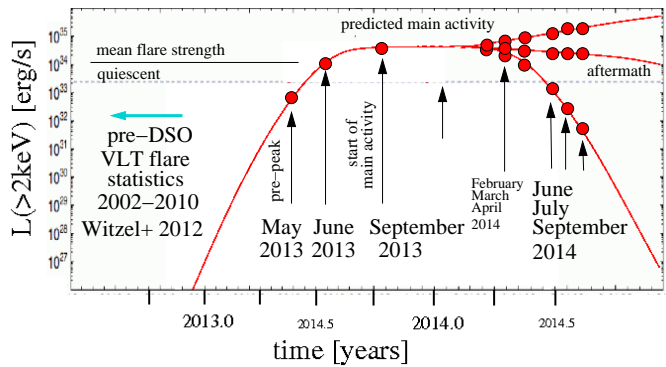}
\caption{The expected SgrA* X-ray luminosity during the fly-by of the DSO (based on Fig.13 in Gillessen et al. 2013b). 
Filled circles indicate times for which we obtained NACO or SINFONI data. 
The threefold branching towards the right indicates three different time 
evolutions of the activity during the aftermath (increase, continuation and drop). 
No exceptional activity has been observed until August 2014.
The source remained in a state of normal activity 
(as during the last years, e.g., Witzel et al. 2012)
with a few flare events that reached the known mean flare level.
} \label{activity}
\end{center}
\end{figure}

In Valencia-S. et al. (2015) and Eckart et al. (2014) 
and announced in the Astronomical Telegram 2014 (ATel \#6285)
we report new near-infrared (1.45$\mu$m - 2.45$\mu$m) observations of the 
Dusty S-cluster Object (DSO/G2) during its approach to the black hole at the 
center of the Galaxy, that were carried out with ESO VLT/SINFONI between February and September 2014
(Fig.~\ref{activity}).
We detect spatially compact Brackett-$\gamma$ and Paschen-$\alpha$ line emission from the 
DSO/G2 at about 30-40~mas east of SgrA*. This is in agreement with the position reported 
by Ghez et al. (2014; ATel \#6110) and Witzel et al. (2014) based on L'-band observations in March 2014. 
The velocity of the source, measured from the red-shifted emission, is about 2700 km/s. 
No strong blue-shifted emission above the noise level is detected at the position of SgrA* or 
any position upstream the presumed orbit. 
The full width at half maximum of the Brackett-$\gamma$ line is ~50 \AA, 
i.e., no significant line broadening with respect to 2013 is observed. 
This is a further indication for the compactness of the source.
For the moment, the flaring activity of the black hole in the near-infrared regime 
has not shown any statistically significant increment.
We conclude that the DSO/G2 source had not yet reached its peribothron before May 2014 and that the 
increased accretion activity of SgrA* is still upcoming.

Through Jalali et al. (2014) we have shown 
how young and dusty stellar objects can be formed in the immediate vicinity of a super-massive black hole.
The observational data was also used to derive the orbit of this object and to predict its peribothron transition.
Due to the presumably high ellipticity of the orbit only very weakly curved sections of the orbit were available 
and first predictions of the peribothron transition time in 2013 (Gillessen et al. 2012) proved to be incorrect.
The inclusion of (or even restrictions to) the Br$\gamma$ line emission resulted in new predictions for
very early 2014.
The fact that in the L'-band the telescope point spread function (PSF) is intrinsically larger and therefore
more susceptible to diffuse extended emission is, probably the main reason for this discrepancy.

\begin{figure}[!ht]
\begin{center}
\includegraphics[width=13cm]{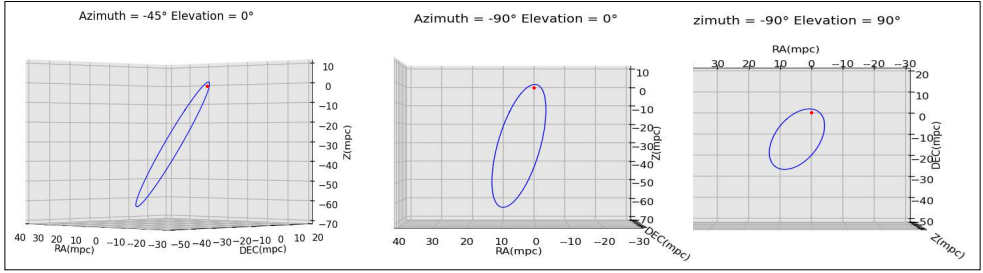}
\caption{Three representations of the DSO orbit around the super-massive black hole shown under different
viewing angels. The orbit calculations include our April-September SINFONI data points. The position of
SgrA* is shown as a small red filled circle. At a distance of 8~kpc one millipasec (mpc) is about 39~mas.
The right panel shows the 'true' sky-view (z-axis towards observed).
} \label{orbit}
\end{center}
\end{figure}

However, the predicted interactions of the gas and dust with the strong gravitational field of SgrA*
have shown that the gas itself may also not be a good probe of the exact orbital motion.
This is supported by the spatial extent and the velocity gradient across the Br$\gamma$ line emission.
It is also highlighted by the expected interaction of the DSO with the ambient medium and the gravitational field.
Therefore, even though the recently derived Br$\gamma$ based orbital solutions are in reasonable agreement
(Meyer et al. 2014), the orbital elements may still be uncertain.
Using the results of our measurements with SINFONI in 2014 and the published Keck data (Meyer et al. 2014)
we revisited the determination of the DSO/G2 orbit. Given the red emission is only about 40~mas East of SgrA* and 
at a radial velocity of about 2700 km/s and only a blue-shifted line emission at about $-$3200m/s was measured after May 2014, 
we obtained a new orbital solution which places the peribothron passage
in mid-May 2014, rather than in 2014.2 as estimated earlier by Meyer et al. (2014). 
However, the orbital elements are very similar to the ones derived earlier. With the high ellipticity and half-axis 
length around  30~mpc we obtain a peribothron distance of the order of 120~AU (Fig.~\ref{orbit}).

\subsection{Emission from Radio to X-rays}

Eckart et al. (2012) model the mm- to X-ray spectrum of the SgrA* flare emission with a 
Synchrotron Self Compton (SSC) mechanism.
They show that the flare emission can be described via a combination of a SSC model 
followed by an adiabatic expansion of the source components (e.g. Eckart et al. 2008).
This involves up-scattered sub-millimeter photons from a compact source component that 
has a spectrum which at the start of the flare peaks at frequencies between several 100 GHz and 2 THz. 
The overall radio-spectrum peaks close to 300 GHz.
Details of the sub-millimeter emission of SgrA* and the surrounding circum nuclear disk is given in
Garcia-Marin et al. (2011).
A combined SSC and adiabatic expansion model can easily explain the observed flare fluxes 
and time delays that cover the spectral range from X-rays to the mm-radio domain. 
So far in the X-ray observable $\ge2$~keV band no elevated continuum flux density level or 
extraordinary X-ray variability - possibly triggered by the DSO fly-by - has been reported 
(Haggard et al. 2014; ATel \#6242)
Such an extra emission would have been expected to originate from the shock-heated gas (Gillessen et al. 2012).
Although SgrA* is extremely faint in the X-ray regime, it is strongly variable in this domain of the electromagnetic spectrum
(Baganoff et al. 2001, 2003, Porquet et al. 2003, 2008, Eckart et al. 2012,
Nowak et al. 2012, Barrire et al.  2014, Mossoux et al. 2015, Neilsen et al.  2013).
The statistical investigation of the near-infrared variability by Witzel et al. (2012)
suggests that the past strong X-ray variations
are potentially linked with the origin of the observed X-ray echos
from a putative strong SgrA* flare
(Revnivtsev et al. 2004, Sunyaev \& Churazov 1998, Terrier et al. 2010, Capelli et al. 2012).

If the underlying mechanism for the NIR/X-ray continuum spectrum of the SgrA* flare emission is
indeed a combination of synchrotron radiation and X-ray radiation through  the Synchrotron Self Compton (SSC) process
then the bright required X-ray flare fluxes can be explained as a natural phenomenon and no exceptional 
variability event needs to be claimed.
All these phenomena make SgrA* an ideal source to study the physics of extremely low
luminosity super massive black holes.
Events like the fly-by of the DSO may dominate the variability of galactic nuclei in this phase 
across the entire electromagnetic spectrum.
However, the onset of the activity phase of the magnetar PSR~J1745-2900 at a separation of only 
about 3 arcseconds from the Galactic center presented a problem 
for the SgrA* monitoring program in 2013 (Mori et al. 2013, Shannon \& Johnston 2013, Rea et al., 2013)
and required higher angular resolution to discern the sources.

\subsection{Infrared Polarimetry}

Linear polarization at 2.2$\mu$m and its statistics and time variation
constrain the physical conditions of the accretion process onto this super-massive black hole.
With an exponent $dN/dS$$\sim$4 of the slope of the number density histogram for flare fluxes 
above 5mJy (see definition in Shahzamanian et al. 2015 and Witzel et al. 2012), the
distribution of polarized flux density is closely linked to the single state power-law distribution of the total
$K_\mathrm{s}$-band flux densities reported earlier.
Closely following the concepts we laid out in Witzel et al. (2011)
we find typical polarization degrees of the order of 10\% to 20\% and
a preferred polarization angle of 13$^\circ$$\pm$15$^\circ$.
Simulations show the uncertainties are probably dominated by observational effects, implying
independently of flux density, that the intrinsic
polarization degree and angle are rather well constrained.
Since the emission is due to optically
thin synchrotron radiation, this preferred polarization angle is very likely coupled to the intrinsic orientation of the
SgrA* system, i.e. a disk or jet/wind scenario associated with the super-massive black hole.
If they are indeed linked to the structural features of the
source the data imply a rather stable geometry and accretion process for the SgrA* system.


\section{Discussion and conclusions}

As we did not detect a strong blue-shifted side of the Br$\gamma$ emission - at least in the predicted strength, 
a prime working hypothesis is that the DSO is a young accreting star. There may be 
an extension to its dust shell and the current estimates of the orbit (assuming
a bound elliptical orbit) bring it quite close to the black hole such that 
we cannot exclude that it may come to an increased activity of SgrA*.
With respect to previous investigations of short and long term variability of SgrA*
no deviation from `normal' activity has been observed until now
(Witzel et al. 2012; Zamaninasab et al. 2010, 2011; Eckart et al. 2006).
Excess of short time flares (with an orbital time scale of 20min or shorter) would have 
allowed us to study accretion disk physics, turbulences 
in a disk or highly fractal accretion stream.
Further spectroscopic measurements of the DSO are required to track it along the
high acceleration part of its orbit in order to get a proper determination of its 
actual orbit.
In the beginning of 2013 the L'-band identification of the DSO got confused. 
Therefore, it is essential to check the L'-band identification after peribothron
to answer the question: 
Does the source simply move out of confusion or has the dust interacted with the 
ISM (enhanced density and radiation field) near the central black hole?

Young accreting pre-main sequence stars are usually variable. The line emission  currently 
shows no strong ($>$30\%) variability over the past years - it is, however, essential
to monitor the strength of the line emission since variability may contribute to the
identification of the object as a young star.
It is currently unclear what the role of the so called `tail emission' of the 
G2/DSO source is. It can best be seen in its Br$\gamma$ line emission. A future determination of
its motion (line of sight and proper motion) will help to decide if it accelerates,
and potentially belongs to the DSO, or whether it belongs to the general back- and foreground emission 
(potentially associated with the mini-spiral) that is abundant and wide spread in the central
parsec of the Milky Way.
Our polarization statistics show that SgrA* must be a very stable system - both in terms
of geometrical orientation of a jet or accretion disk and in terms of the variability spectrum 
which must be linked to the accretion rate.
Hence, polarization and variability measurements are the ideal tool to probe for any change in the system as a 
function of the DSO fly-by (Shahzamanian et al. 2015).
In order to make progress in understanding the physical processes responsible for the radio to X-ray emission of SgrA*,
and to better discern the flaring non-thermal from the extended non-variable Bremsstrahlung
component (Baganoff et al. 2003) it is imperative for future X-ray missions to provide higher angular resolution and sensitivity.
This will then allow us to expand our studies to the X-ray flux density variations in faint phases.
The possible counterparts of these fainter X-ray variations could the be studied in the NIR and radio domain.
Higher sensitivity in the X-ray domain may also open the door to search and potentially find line emission
from the immediate vicinity of SgrA*. Alternatively, this could be achieved for particularly bright X-ray flares as well.
X-ray line emission could be used as a further tool to study the dynamics of matter (e.g., Valencia-S. et al. 2012) 
near the super-massive black hole.
\\
\\
{\bf Acknowledgments:}\,\,\, 
This work was supported in part by the Deutsche Forschungsgemeinschaft
(DFG) via the Cologne Bonn Graduate School (BCGS),
the Max Planck Society through
the International Max Planck Research School (IMPRS) for Astronomy and
Astrophysics, as well as
special funds through the University of Cologne.
M. Zajacek, S. Smajic, B. Shahzamanian, N. Sabha, M. Parsa, and A. Borkar are members of the IMPRS.
The research leading to these results received funding from the
European Union Seventh Framework Program (FP7/2007-2013) under grant
agreement n312789.
Part of this work was supported by fruitful discussions
with members of the European Union funded COST Action MP0905: Black
Holes in a violent Universe and
the Czech Science Foundation -- DFG collaboration (No. 13-00070J).
This work was co-funded under the Marie Curie Actions of the European
Commission (FP7-COFUND).
M. Garc\'{\i}a-Mar\'{\i}n is supported by the German
federal department for education and research (BMBF) under
the project number 50OS1101.
We are grateful to all members of the ESO PARANAL team.

\end{document}